\newif\iffigs\figsfalse              
\figstrue                            
\newif\ifbbB\bbBfalse                
\bbBtrue                             


\input harvmac
\overfullrule=0pt
\def\Title#1#2{\rightline{#1}
 \ifx\answ\bigans
  \nopagenumbers\pageno0\vskip1in\baselineskip15pt plus1pt minus1pt
 \else
  \def\listrefs{\footatend\vskip 1in\immediate\closeout\rfile\writestoppt
   \baselineskip=14pt\centerline{{\bf References}}\bigskip{\frenchspacing
   \parindent=20pt\escapechar=` \input refs.tmp\vfill\eject}\nonfrenchspacing}
  \pageno1\vskip.8in\fi 
\centerline{\titlefont #2}\vskip .5in}

\ifx\answ\bigans
 \def\tcbreak#1{}
\else
 \def\tcbreak#1{\cr&{#1}}
\fi
\newcount\figno \figno=0
\iffigs
 \message{If you do not have epsf.tex (to include figures),}
 \message{change the option at the top of the tex file.}
 \input epsf
 \def\fig#1#2#3{\par\begingroup\parindent=0pt\leftskip=1cm\rightskip=1cm
  \parindent=0pt \baselineskip=11pt \global\advance\figno by 1 \midinsert
  \epsfxsize=#3 \centerline{\epsfbox{#2}} \vskip 12pt
  {\bf Fig. \the\figno:} #1\par \endinsert\endgroup\par }
\else
 \message{No figures will be included. See TeX file for more information.}
 \def\fig#1#2#3{\global\advance\figno by 1 \vskip .5in
  \centerline{\bf Figure \the\figno} \vskip .5in}
\fi
\ifbbB
 \message{If you do not have msbm (blackboard bold) fonts,}
 \message{change the option at the top of the tex file.}
 \font\blackboard=msbm10 
 \font\blackboards=msbm7 \font\blackboardss=msbm5
 \newfam\black \textfont\black=\blackboard
 \scriptfont\black=\blackboards \scriptscriptfont\black=\blackboardss
 \def\Bbb#1{{\fam\black\relax#1}}
\else
 \def\Bbb{\bf}
\fi

\def\NPB#1#2#3{{\sl Nucl. Phys.} \underbar{B#1} (#2) #3}
\def\PLB#1#2#3{{\sl Phys. Lett.} \underbar{#1B} (#2) #3}
\def\PRL#1#2#3{{\sl Phys. Rev. Lett.} \underbar{#1} (#2) #3}

\def\undertext#1{$\underline{\smash{\hbox{#1}}}$}
\def\til{\widetilde}
\def\bar{\overline}
\def\vev#1{{\langle #1\rangle}}
\def\bo{\hbox{1\kern -.23em {\rm l}}}
\def\bC{{\Bbb C}}

\def\bS{{\bf S}}
\def\bT{{\bf T}}
\def\bZ{{\Bbb Z}}
\def\CN{{\cal N}}
\def\CO{{\cal O}}
\def\t{{}^t\!}
\def\tQ{{\til Q}}
\def\tq{{\til q}}
\def\tm{{\til m}}
\def\p{\partial}

\def\smdr{\mathbin{\hbox{\hskip2pt\vrule height 5.0pt depth -.3pt
 width .25pt \hskip-2pt$\times$}}}

\Title{hep-th/9509175, RU-95-61, UK/95-14}
{\vbox{\centerline{The Vacuum Structure of $\CN{=}2$ Super--QCD}
\medskip
\centerline{with Classical Gauge Groups}}}
\centerline{Philip C. Argyres}
\smallskip
\centerline{\it Department of Physics and Astronomy}
\centerline{\it Rutgers University}
\centerline{\it Piscataway NJ 08855--0849, USA}
\centerline{\tt argyres@physics.rutgers.edu}
\smallskip
\centerline{and}
\smallskip
\centerline{Alfred D. Shapere}
\smallskip
\centerline{\it Dept.\ of Physics and Astronomy}
\centerline{\it University of Kentucky}
\centerline{\it Lexington, KY 40506--0055, USA}
\centerline{\tt shapere@amoeba.pa.uky.edu}
\baselineskip 18pt
\bigskip
\noindent
{\bf Abstract:} We determine the vacuum structure of $\CN{=}2$
supersymmetric QCD with
fundamental quarks for gauge groups $SO(n)$ and $Sp(2n)$, extending
prior results for $SU(n)$. The solutions are all given in terms of
families of hyperelliptic Riemann surfaces of genus equal to the rank
of the gauge group.  In the scale invariant cases, the solutions all
have exact S-dualities which act on the couplings by subgroups of
$PSL(2,\bZ)$ and on the masses by outer automorphisms of the flavor
symmetry.  They are shown to reproduce the complete pattern of symmetry
breaking on the Coulomb branch and predict the correct weak--coupling
monodromies.  Simple breakings with squark vevs provide further
consistency checks involving strong--coupling physics.
\Date{September 1995}

\nref\SWI{N. Seiberg and E. Witten, ``Electric-Magnetic Duality, 
	Monopole Condensation, and Confinement in N=2 Supersymmetric 
	Yang-Mills Theory,'' hep-th/9407087, \NPB{426}{1994}{19}.}
\nref\SWII{N. Seiberg and E. Witten, ``Monopoles, Duality, and Chiral 
	Symmetry Breaking in N=2 Supersymmetric QCD,'' hep-th/9408099, 
	\NPB{431}{1994}{484}.}
\nref\AF{P.C. Argyres and A.E. Faraggi, ``Vacuum Structure and Spectrum
	of N=2 Supersymmetric SU(n) Gauge Theory,'' hep-th/9411057, 
	\PRL{74}{1995}{3931}.}
\nref\KLTY{A. Klemm, W. Lerche, S. Theisen and S. Yankielowicz, ``Simple
	Singularities and N=2 Supersymmetric Yang-Mills Theory,'' 
	hep-th/9411048, \PLB{344}{1995}{169}.}
\nref\APS{P.C. Argyres, M.R. Plesser, and A.D. Shapere, ``The Coulomb
	Phase of N=2 Supersymmetric QCD,'' hep-th/9505100, 
	\PRL{75}{1995}{1699}.}
\nref\HO{A. Hanany and Y. Oz, ``On the Quantum Moduli Space of N=2
	Supersymmetric SU(N) Gauge Theories,'' hep-th/9505075.}
\nref\DS{M.R. Douglas and S.H. Shenker, ``Dynamics of SU(N) Supersymmetric 
	Gauge Theory,'' hep-th/9503163, \NPB{447}{1995}{271}.}
\nref\AD{P.C. Argyres and M.R. Douglas, ``New Phenomena in SU(3) 
	Supersymmetric Gauge Theory,'' hep-th/9505062, \NPB{448}{1995}{93}.}
\nref\MN{J. Minahan and D. Nemeschansky, ``Hyperelliptic Curves for
	Supersymmetric Yang-Mills,'' hep-th/9507032.}
\nref\UDBS{U.H. Danielsson and B. Sundborg, ``The Moduli Space and 
	Monodromies of N=2 Supersymmetric SO(2r+1) Yang-Mills Theory,'' 
	hep-th/9504102.}
\nref\BL{A. Brandhuber and K. Landsteiner, ``On the Monodromies of N=2
	Supersymmetric Yang-Mills Theory with Gauge Group SO(2n),'' 
	hep-th/9507008.}
\nref\GKMMM{A. Gorskii, I. Krichever, A. Marshakov, A. Mironov, and A.
	Morozov, ``Integrability and Seiberg-Witten Exact Solution,''
	hep-th/9505035, \PLB{355}{1995}{466}.}
\nref\MW{E. Martinec and N. Warner, ``Integrable Systems and Supersymmetric
	Gauge Theory,'' hep-th/9509161.}
\nref\NT{T. Nakatsu and K. Takasaki, ``Whitham-Toda Hierarchy and N=2
	Supersymmetric Yang-Mills Theory,'' hep-th/9509162.}
\nref\IS{K. Intriligator and N. Seiberg, ``Duality, Monopoles, Dyons, 
	Confinement, and Oblique Confinement in Supersymmetric $SO(N_c)$ 
	Gauge Theories,'' hep-th/9503179, \NPB{444}{1995}{125}.}
\nref\nrt{N. Seiberg, private communication.}
\nref\MNcorr{J. Minahan and D. Nemeschansky, private communication.}
\nref\KLTYii{A. Klemm, W. Lerche, S. Theisen and S. Yankielowicz, 
	``On the Monodromies of N=2 Supersymmetric Yang-Mills
	Theory,'' hep-th/9412158.}
\nref\dWLVP{B. de Wit, P.G. Lauwers, and A. Van Proeyen, ``Lagrangians of
	N=2 Supergravity-Matter Systems,'' \NPB{255}{1985}{569}.}

\newsec{Introduction and Summary}

Seiberg and Witten \refs{\SWI,\SWII} found the exact vacuum structure
and spectrum of four-dimensional $\CN{=}2$ supersymmetric $SU(2)$ QCD.
Their analysis was extended to gauge group $SU(r{+}1)$ with matter in
the fundamental representation \refs{\AF, \KLTY, \APS, \HO}.  In this
paper we further extend this analysis to include the simple gauge groups
$Sp(2r)$ with matter in the fundamental representation and $SO(n)$
with vector matter.  

Following \APS, we assume the solutions are uniformly described in
terms of the moduli spaces of families of hyperelliptic Riemann
surfaces. We then construct the unique solution consistent with this
assumption by induction on $r$, the rank of the gauge group, using the
patterns of symmetry breaking obtained by condensation of the complex
scalar in the adjoint of the gauge group.

The resulting solutions satisfy a number of restrictive consistency
requirements.  First, they reproduce the complete pattern of symmetry
breaking on the Coulomb branch (only a small subset of these breaking
patterns are used in the induction argument).  Second, there exist
meromorphic one-forms on these curves whose periods generate the
spectrum of low-energy excitations.  These one-forms obey a set of
differential equations and conditions on their residues which, for a
generic curve, need have no solution.  Third, these curves correctly
reproduce all the weak-coupling monodromies in the $Sp(2r,\bZ)$
duality group.  Finally, they reproduce the expected pattern of
symmetry breaking on the Higgs branches.  The latter property is a 
test of consistency at strong coupling. 

The qualitative features of the solutions are similar to those of the
$SU(r{+}1)$ theory:  the generic vacuum is $U(1)^r$ $\CN{=}2$
supersymmetric Abelian gauge theory;  along special submanifolds of the
Coulomb branch there is a rich spectrum of vacua, with massless
electrically and magnetically charged states analogous to those
studied in \DS, new non-trivial fixed points like those 
studied in \AD, as well as nonabelian Coulomb phases;
and, in the cases where the beta function vanishes,
the solutions exhibit exact scale invariance and strong-weak coupling
duality.  The duality acts in all cases as a subgroup of $SL(2,\bZ)$
on the couplings and on the masses by outer automorphisms of the
flavor symmetry.  The specific solutions follow; we first recall
the $SU(r{+}1)$ solution with coupling $\tau$ \APS.  

\undertext{$SU(r{+}1)$ with $N_f=2r{+}2$} fundamental hypermultiplets has
the curve and one-form
		\eqnn\Esucurve
		$$\displaylines{
	y^2  = \prod_{a=1}^{r+1} (x - \phi_a)^2
	+ 4h(h+1) \prod_{j=1}^{2r+2} (x - m_j - 2h \mu) ,\qquad
	\sum \phi_a = 0 ,\cr
	\hfill\lambda  = {x-2h\mu\over2\pi i}d\,\log\left(\prod(x-\phi_a)-y
	\over \prod(x-\phi_a) + y \right),\hfill\llap{\Esucurve}\cr
	h(\tau)  = {\vartheta_2^4 \over \vartheta_4^4-\vartheta_2^4}.\cr
		}$$
The masses $m_j$ transform in the adjoint of the $U(1){\times}SU(N_f)$ 
flavor group, and $\mu \equiv (1/N_f)\sum m_j$ is the flavor-singlet mass. 
For $r{=}1$ this curve is equivalent to the 4-flavor $SU(2)$ curve \SWII.
For $r{>}1$ it is invariant under a $\Gamma^0(2)\subset
PSL(2,\bZ)$ duality group generated by $T^2{:}\ \tau\to\tau{+}2$, and
$S{:}\ \tau\to-1/\tau,\ m_j\to m_j{-}2\mu$.  The solutions for the
asymptotically free theories with $N_f < 2r{+}2$ flavors are obtained
by taking $2r{+}2{-}N_f$ masses ${\sim}M{\to}\infty$, while keeping
$\Lambda^{2r+2-N_f} = q M^{2r+2-N_f}$ finite.  Another
$SU(3)$ curve has been presented in \MN;  it differs from the above
curve only by a (non-perturbative) redefinition of the coupling $\tau$.

\undertext{$Sp(2r)$ with $N_f = 2r{+}2$} fundamental hypermultiplets has 
curve and one-form
		\eqnn\Espcurve $$\displaylines{
	x y^2 = \left(x\prod_{a=1}^r(x - \phi^2_a)
	+ g   \prod_{j=1}^{2r+2} m_j \right)^2
	- g^2 \prod_{j=1}^{2r+2} (x - m_j^2) ,\cr
	\hfill\lambda = {\sqrt x\over 2\pi i}d\,\log\left( 
	{x\prod(x-\phi_a^2) + g\prod m_j - \sqrt x y \over 
	 x\prod(x-\phi_a^2) + g\prod m_j + \sqrt x y} \right) ,
	\hfill\llap{\Espcurve}\cr
	g(\tau) = {\vartheta_2^4 \over \vartheta_3^4+\vartheta_4^4}.
		}$$
(Note that the right side of the curve is divisible by $x$, and that the
$\sqrt x$'s in $\lambda$ cancel upon expanding the derivative.)
The masses transform in the adjoint of the $SO(2N_f)$ flavor group.
For $r{=}1$ this reduces to the 4-flavor $SU(2)$ curve \SWII, which has
a $PSL(2,\bZ)$ duality acting on the coupling and transforming the
masses by the $S_3$ outer automorphisms of the $SO(8)$ flavor symmetry.
For $r{>}1$ this solution is invariant under a $\Gamma_0(2)\subset
PSL(2,\bZ)$ duality group generated by $T{:}\ \tau\to\tau{+}1,\ \prod m_j
\to-\prod m_j$, and $ST^2S{:}\ \tau\to\tau/(1{-}2\tau)$.  The solutions
for the asymptotically free theories with $N_f < 2r{+}2$ flavors are
obtained by taking $2r{+}2{-}N_f$ masses ${\sim}M{\to}\infty$, while keeping
$\Lambda^{2r+2-N_f} = q M^{2r+2-N_f}$ finite.

\undertext{$SO(2r{+}1)$ with $N_f = 2r{-}1$} vector hypermultiplets has 
curve and one-form

		\eqnn\Esooddcurve $$\displaylines{
	y^2 = x\prod_{a=1}^r(x - \phi^2_a)^2
	+ 4f x^2 \prod_{j=1}^{2r-1} (x - m_j^2) ,\cr
	\hfill\lambda = {\sqrt x\over 2\pi i}d\,\log\left( 
	{x\prod(x-\phi_a^2) - \sqrt x y \over 
	 x\prod(x-\phi_a^2) + \sqrt x y} \right) ,
	\hfill\llap{\Esooddcurve}\cr
	f(\tau) = {\vartheta_2^4\vartheta_4^4 \over 
	(\vartheta_2^4-\vartheta_4^4)^2}.
		}$$
The masses transform in the adjoint of the $Sp(2N_f)$ flavor group.
For $r{=}1$ this reduces to the $SU(2)$ curve with one adjoint 
hypermultiplet \SWII, which has a $PSL(2,\bZ)$ duality acting on the 
coupling.  For $r{>}1$ this solution is invariant under a $\Gamma^0(2)
\subset PSL(2,\bZ)$ duality group generated by $T^2{:}\ \tau\to\tau{+}2$,
and $S{:}\ \tau\to-1/\tau$.  
The solutions for the asymptotically free theories with $N_f<2r{-}1$ 
flavors are obtained by taking $2r{-}1{-}N_f$ masses ${\sim}M{\to}\infty$ 
keeping $\Lambda^{4r-2-2N_f} = q M^{4r-2-2N_f}$ finite.  In the 
Yang-Mills case ($N_f{=}0$) this solution is equivalent to the solution 
of \UDBS.

\undertext{$SO(2r)$ with $N_f = 2r{-}2$} vector hypermultiplets has the curve
		\eqn\Esoevencurve{
	y^2 = x\prod_{a=1}^r(x - \phi^2_a)^2
	+ 4f x^3 \prod_{j=1}^{2r-2} (x - m_j^2),
		}
with one-form and $f(\tau)$ the same as for the $SO(2r{+}1)$ case
\Esooddcurve, and the same duality group as well.  In the Yang-Mills
case this solution is equivalent to the solution of \BL.

The coupling dependence of each of these solutions is given in terms
of the usual Jacobi theta functions defined by\footnote*{Note that
these theta functions are labelled differently from the $\theta_i$
used in \refs{\SWII, \APS}.  The relation between the two is $\theta_1
{=} \vartheta_2$, $\theta_2 {=} \vartheta_4$, $\theta_3 {=} \vartheta_3$.}
               \eqn\thetafnc{\eqalign{
        \vartheta_2(\tau) &= \sum_{n\in\bZ} q^{(n+1/2)^2} ,\qquad\qquad
        \vartheta_2^4 = \ \ \ \, 16q+\CO(q^3),\cr
        \vartheta_3(\tau) &= \sum_{n\in\bZ} q^{n^2} ,\qquad\qquad\qquad
        \ \vartheta_3^4 = 1+8q+\CO(q^2),\cr
        \vartheta_4(\tau) &= \sum_{n\in\bZ} (-1)^n q^{n^2} ,\qquad\qquad
        \vartheta_4^4 = 1-8q+\CO(q^2),\cr
                }}
which satisfy the Jacobi identity $\vartheta_2^4{-}\vartheta_3^4{+}
\vartheta_4 ^4 = 0$.  Here 
		\eqn\deftau{
	q \equiv e^{i\pi\tau} \equiv e^{i\theta}e^{-8\pi^2/g^2} ,
		}
where $\theta$ is the theta angle and $g$ the gauge coupling.
Under the modular transformations $S{:}\ \tau \to -1/\tau$ and $T{:}\ 
\tau \to \tau{+}1$, the theta functions transform as
                \eqn\thetaST{\eqalign{
        \vartheta_2^4(-1/\tau) &= -\tau^2\vartheta_4^4(\tau) ,\cr
        \vartheta_3^4(-1/\tau) &= -\tau^2\vartheta_3^4(\tau) ,\cr
        \vartheta_4^4(-1/\tau) &= -\tau^2\vartheta_2^4(\tau) ,\cr
                }\qquad\qquad\qquad \eqalign{
        \vartheta_2^4(\tau+1) &= -\vartheta_2^4(\tau) ,\cr
        \vartheta_3^4(\tau+1) &= \vartheta_4^4(\tau) ,\cr
        \vartheta_4^4(\tau+1) &= \vartheta_3^4(\tau) .\cr
                }}

Two natural extensions of our work would be to apply similar
techniques to other groups and other matter representations. Regarding
the latter, we can write down solutions in a number of special cases,
using the equivalences between low--rank groups. These are $SO(3)$
with adjoint matter, $SO(5)$, $SO(6)$, and $SO(8)$ with spinor matter,
$Sp(4)$ with ${\bf 5}$'s, $SU(4)$ with {\bf 6}'s, and $SU(2)\times
SU(2)$ with ({\bf 2},{\bf 2}) matter.  These examples should provide
useful initial matching conditions for inductive generalizations to
higher--rank groups.

Regarding the possibility of extending our results to include all Lie
groups, it seems unlikely that the curves for the exceptional groups
will be hyperelliptic.  The hyperelliptic {\it Ansatz} is essentially
the simplest assumption we can make about the form of the surface;
although it works for the theories analyzed below, we do not know of
any physical argument indicating it should be true of other theories
as well.  One of the main technical reasons it works for the classical
groups is that the basis of holomorphic one-forms on a genus-$r$
hyperelliptic surface is $\omega_\ell = x^{r-\ell}dx/y$ for
$\ell{=}1,\ldots,r$.  When $x$ and $y$ are assigned definite
dimensions (or R-charges), the set of $\omega_\ell$ have evenly spaced
dimensions.  There is a natural one-form solving the differential
equation $\p\lambda/\p s_\ell
\propto \omega_\ell$ if the basis of polynomial invariants of the group
$s_\ell$, $\ell{=}1,\dots,r$, also has evenly spaced dimensions.  This
is indeed the case for the classical groups.\footnote*{Although
$SO(2r)$ has an ``extra'' invariant $t{=}\sqrt s_r$, dividing by a
global $\bZ_2$ symmetry corresponding to the outer automorphism of
$SO(2r)$ that takes $t \to -t$ allows us to consider a curve depending
only on $s_r$; presumably, a curve describing the $SO(2r)$ theory
without dividing by this symmetry would not be hyperelliptic.}
Products of simple groups suffer from the same problem.  Recent papers
\refs{\GKMMM,\MW,\NT} on the Yang--Mills curves have proposed a uniform
framework for all gauge groups.  It would be very interesting to
understand the QCD solutions presented here in terms of this framework,
and thus dispense of the need for additional assumptions on the form
of the solution.

At least as interesting as the extension of the results of this paper
to other groups and representations would be the extraction of new
physics from them. For example, there is reason to believe that the
$SO(n)$ theories possess a richer phenomenological structure than the
$SU(n)$ theories, based on the $\CN{=}1$ results of \IS. A clearer
understanding of the origin of such $\CN{=}1$ phenomena might be
gained within the framework of our $\CN{=}2$ solutions.

In the remainder of this paper we briefly review $U(1)^r$ duality in
the low--energy $\CN{=}2$ supersymmetric effective theory, describe
our basic {\it Ansatz} for the solution, and then proceed to derive
the curves and one-forms for the $Sp(2r)$, $SO(2r{+}1)$, and $SO(2r)$
theories in turn.  In each case the consistency requirements on the
Coulomb branch mentioned above are checked. Following our discussion
of the various Coulomb branches, we check a Higgs branch consistency
requirement relating the different solutions.

\newsec{U(1)${}^r$ Duality and the Hyperelliptic Ansatz}

$\CN{=}2$ QCD is described in terms of $\CN{=}1$ superfields by a
chiral field strength multiplet $W$ and a chiral multiplet $\Phi$ both
in the adjoint of the gauge group, together with chiral multiplets
$Q^j$ in a representation $R$, and $\tQ_j$ in the complex conjugate
representation $\bar R$ of the gauge group.  The flavor index $j$ runs
from $1$ to $N_f$.  The Lagrangian contains $\CN{=}1$ gauge--invariant
kinetic terms for the fields with gauge coupling constant $\tau$ and
superpotential ${\cal W} = \sqrt2 \tQ_j \Phi Q^j$.  Classically, the
global symmetries are the $U(1) {\times} SU(2)$ chiral R-symmetry, any
outer automorphisms of the gauge group which leave the representation
$R{\oplus}\bar R$ invariant, and the flavor symmetry, a subgroup of
$U(2N_f)$ determined by the superpotential interaction.  The $\CN{=}2$
invariant quark mass matrix $M$ is a complex matrix in the adjoint of
the flavor group satisfying $[M,M^\dagger]{=}0$, implying that $M$ can
be taken to be in the Cartan subalgebra of the flavor group by a
flavor rotation.  In the quantum theory the $U(1)$ R-symmetry is
generally broken by anomalies to a discrete subgroup.

The theory has a rich vacuum structure consisting of Higgs, Coulomb,
and mixed branches.  We focus on the Coulomb branch since a
nonrenormalization theorem \nrt\ implies that only the Coulomb branch
can receive quantum corrections; the Higgs branch is determined by the
classical equations of motion alone.  We will use this fact in Section
7 to find relations between the solutions for various simple gauge
groups.  On the Coulomb branch the vevs of the lowest components of the
chiral superfields satisfy $q^j{=}\tq_j{=}0$ and $[\phi, \phi^\dagger]
{=} 0$.  This implies that $\phi$ can be chosen by a color rotation to
lie in the complexified Cartan subalgebra of the gauge group.
$\vev\phi$ generically breaks the gauge symmetry to $U(1)^r$, where $r$
is the rank of the gauge group, and gives all the quarks masses, so the
low energy effective theory is an $\CN{=}2$ supersymmetric $U(1)^r$
Abelian gauge theory.  Classically, for special values of $\vev\phi$
and the quark masses $m_j$, the unbroken gauge group will include
nonabelian factors or massless quarks.

Assuming $\CN{=}2$ supersymmetry is not dynamically broken, the Coulomb
vacua are not lifted by quantum effects.  At a generic point, the low
energy effective Lagrangian can be written in terms of $\CN{=}2$\
\ $U(1)$ gauge multiplets $(A_\mu,W_\mu)$, where $\mu,\nu=1,\ldots,r$
and label quantities associated to each of the $U(1)$ factors.  We
denote the scalar component of $A_\mu$ by $a_\mu$, which we will also
take to stand for its vev.  The effective Lagrangian is determined by
an analytic prepotential ${\cal F}(A_\mu)$,
		\eqn\effL{
{\cal L}_{\rm eff} = {\rm Im}{1\over4\pi} \left[ \int\!\! d^4\theta\,
A_D^\mu\,\bar{A_\mu} + {1\over2}\int\!\!d^2\theta\,\tau^{\mu\nu}\,W_\mu
W_\nu \right] ,
		}
with dual chiral fields $A_D^\mu \equiv \p{\cal F}/\p A_\mu$, and
effective couplings $\tau^{\mu\nu} \equiv \p^2{\cal F} /\p A_\mu\p
A_\nu$.  Near submanifolds of moduli space where extra states become
massless the range of validity of \effL\ shrinks to zero; on these
singular submanifolds the effective Lagrangian must be replaced with
one which includes the new massless degrees of freedom.

The $U(1)^r$ theory has a lattice of allowed electric and magnetic
charges, $q^\mu$ and $h_\mu$.  Generically, the bare masses break the
flavor symmetry to $U(1)^{N_f}$, so states have associated quark number
charges $n^j\in\bZ$.  A BPS saturated $\CN{=}2$ multiplet with quantum
numbers $q^\mu$, $h_\mu$, and $n^j$ has a mass given by \SWII\ $M =
|a_\mu q^\mu {+} a^\mu_D h_\mu {+} m_j n^j|$.  The physics described by
the $U(1)^r$ effective theory is invariant under duality
transformations $(\bS,\bT) \in Sp(2r;\bZ) \smdr \bZ^{N_f}$ which act on
the fields and charges as ${\bf a} \to \bS{\cdot}{\bf a}+ \bT {\cdot}
{\bf m}$,\ \ ${\bf h} \to \t\bS^{-1}{\cdot}{\bf h}$, and ${\bf n} \to
-\bT{\cdot}{\bf h} + {\bf n}$.  Here we have defined the column vectors
${\bf a} \equiv \t(a_D^\mu, a_\nu)$, ${\bf m} \equiv (m_j)$, ${\bf
h} \equiv \t(h_\mu,q^\nu)$, and ${\bf n} \equiv (n^j)$.  Encircling a
singular submanifold in moduli space may produce a non-trivial duality
transformation.  The monodromy around a submanifold where one dyon with
charges $({\bf h},{\bf n})$ is massless is
		\eqn\dyonmon{
\bS = \bo+{\bf h}\otimes\t({\bf J}\cdot{\bf h}), \qquad \bT = {\bf
n}\otimes\t({\bf J}\cdot{\bf h}),
		}
where ${\bf J}={\ 0\ 1\choose -1\ 0}$ is the symplectic metric.
The action of $\bT$ on the periods corresponds to the freedom to shift
the global quark--number current by a multiple of a $U(1)$ gauge
current \SWII.

Our aim is to determine the analytic prepotential ${\cal F}$ of the
low--energy Abelian theory everywhere on moduli space.  Let $\{s_\ell\}$
be good coordinates on the Coulomb branch.  We will assume, following
\refs{\SWI,\SWII,\AF,\KLTY}, that the effective coupling
$\tau^{\mu\nu}(s_\ell)$ is the period matrix of a genus $r$ Riemann
surface $\Sigma(s_\ell)$ varying holomorphically over moduli space, and
the vevs of the scalar fields and their duals are given by $a_D^\mu =
\oint_{\alpha^\mu} \lambda$ and $a_\nu = \oint_{\beta_\nu} \lambda$,
where $\lambda (s_\ell)$ is a meromorphic form on $\Sigma(s_\ell)$.
Here $(\alpha^\mu,\beta_\nu)$ is a basis of $2r$ one-cycles on $\Sigma$
with the standard intersection form $\vev{\alpha^\mu,\beta_\nu} =
\delta^\mu_\nu$, $\vev{\alpha^\mu,\alpha^\nu} =
\vev{\beta_\mu,\beta_\nu}=0$.  The residues at the poles of $\lambda$
must be integral linear combinations of $m_j$, the bare quark masses,
and the form must satisfy
		\eqn\difeq{
{\p\lambda\over\p s_\ell} = \omega_\ell + d f_\ell,
		}
with $\omega_\ell$ a basis of $r$ holomorphic one-forms on $\Sigma$,
and $f_\ell$ arbitrary functions.  Duality
transformations $(\bS,\bT)$ have the effect of redefining 
the symplectic basis  and shifting the winding numbers
of the cycles around each of the poles according to \dyonmon.  
The condition on the residues of $\lambda$ guarantees that the correct
action of $\bT$ on the vevs is realized.

We will further assume, as in \refs{\AF,\KLTY,\APS}, that $\Sigma$ is
a hyperelliptic Riemann surface with polynomial dependence on the
coordinates $s_\ell$ and the masses $m_j$. A curve $y^2= \wp(x)$,
where $\wp(x)$ is a polynomial in $x$ of degree $2r{+}2$, describes a
hyperelliptic Riemann surface of genus $r$ as a double--sheeted cover
of the $x$-plane branched over $2r{+}2$ points.

\newsec{Sp(2r)}

\subsec{Symmetries}

The unitary symplectic group $Sp(2r)$ has rank $r$ and dimension
$2r^2{+}2$.  The adjoint representation has index $2r{+}2$, and the
pseudoreal fundamental representation has dimension $2r$ and index
$1$.  Thus the $\CN{=}2$ beta function for the theory with $N_f$
fundamental hypermultiplets is $i\pi\beta = N_f{-}2r{-}2$ and the
instanton factor is $\Lambda^{2r+2-N_f}$ in the asymptotically free
cases.

On the Coulomb branch, the adjoint chiral superfield $\Phi$ has
expectation values that can be diagonalized as $\vev\phi = {\rm
diag}(\phi_1, \ldots, \phi_r, -\phi_1, \ldots, -\phi_r)$.  The
gauge--invariant combinations of the $\phi_a$'s are all the symmetric
polynomials $s_\ell$ in $\phi_a^2$ up to degree $2r$.  These are
generated by
		\eqn\geninvsp{
\sum_{\ell=0}^r s_\ell x^{r-\ell} = \prod_{a=1}^r (x-\phi_a^2) .
		}
The general (maximal) adjoint breaking is $Sp(2r) \to Sp(2r{-}2k)
{\times} SU(k) {\times} U(1)$, which occurs via an expectation value
$\vev\phi=(M,\ldots,M,0,\ldots,0)$ with $r{-}k$ $M$'s.  The fundamental
decomposes as $\bf{2r} = ({\bf 2r{-}2k}, \bf1) \oplus (\bf1,{\bf 2k})
\oplus (\bf1,{\bf \overline{2k{+}1}})$ under $Sp(2r{-}2k) {\times}
SU(k)$.

The flavor symmetry of this theory is $O(2N_f)$, and an $\CN{=}2$
supersymmetric mass term can be skew--diagonalized to masses $\pm m_j$,
$j{=}1,\ldots,N_f$.  The $O(2N_f)$ invariants in terms of these mass
eigenvalues are all the symmetric polynomials in the $m_j^2$ up to
degree $2N_f{-}2$, plus the product of the masses $\prod_j^{N_f} m_j$.
To see the $O(2N_f)$ flavor symmetry, define the $2N_f$-component quark
$X^{\hat\jmath}_a=(Q_a^j {+} i\tQ^j_a, Q_a^j {-} i\tQ^j_a)$ so the kinetic 
terms are ${\cal K} \propto X_a{\cdot}(X^+)^a$ and the superpotential 
is ${\cal W} = X_a{\cdot}\t X_b\,\Phi^{ab}$ since $\Phi^{ab}$ is symmetric.  
(The adjoint of $Sp(2r)$ is the symmetric product of two fundamentals.)
The global symmetry of $\cal K$ is $U(2N_f)$, while $\cal W$ is left 
invariant by $O(2N_f,\bC)$.  Their intersection is $O(2N_f)$.

$O(2N_f)$ differs from $SO(2N_f)$ by a global $\bZ_2$ acting by
interchanging $Q^1 \leftrightarrow \til Q^1$, leaving the other quarks
invariant, which can also be thought of as the action of the
outer automorphism of $SO(2N_f)$.  This classical $\bZ_2$ is
anomalous.

\subsec{Curve and One-Form}

We now determine the form of the curve for the $Sp(2r)$ theory by
imposing the following requirements: (1) that the form of the curve be
uniform in $r {\ge} 1$, and (2) that the curve have the form
		\eqn\Ecrvform{
y^2 = P_{cl}(x,\phi_a) + Q_{qu}(x,\phi_a,m_i,\Lambda),
		}
where the ``quantum piece'' $Q_{qu}$ vanishes when $\Lambda {\to} 0$
or, in the scale invariant case, $\tau \to +i\infty$.  Only gauge- and
flavor--invariant combinations of the $\phi_a$ and $m_j$ should appear
in \Ecrvform.

In the weak coupling limit $\Lambda {\to} 0$ the branch points of this
curve are at the zeros of $P_{cl}$.  When two (or more) of these
coincide one or more cycles of the Riemann surface degenerate,
corresponding to some charged state becoming massless.  Such massless
states appear at weak coupling whenever two of the $\phi_a$ coincide
(corresponding to an unbroken $SU(2)$ gauge subgroup of $Sp(2r)$) or
one of the $\phi_a$ vanishes (corresponding to an unbroken $Sp(2)$
gauge group).  The symmetry in the $\phi_a$ together with the fact
that the whole curve must be singular as $\Lambda {\to} 0$ implies
$P_{cl}$ must be of the form
		\eqn\Eclguessii{
P_{cl} = x^\epsilon \prod_a^r (x -\phi_a^2)^2 \equiv x^\epsilon P^2,
\qquad \epsilon = 1 \ {\rm or} \ 2.
		}
This has the right degree to describe a curve of genus $r$ if $\epsilon
{=}2$; if $\epsilon {=} 1$ it also describes a genus $r$ surface, but
with one of the branch points fixed at infinity on the $x$-plane.  Note
that $x$ has dimension two.  The known solution \refs{\SWI,\SWII} for
the $SU(2) {\simeq} Sp(2)$ case has one branch point fixed at infinity,
so, in order to have a description uniform in $r$, we choose $\epsilon
{=} 1$.

To determine the ``quantum'' piece of the curve it is convenient to
consider the most general possible form of the curve for $Sp(2r{+}2k)$
with $2r{+}2$ flavors, and to consider the resulting curve upon
breaking to $Sp(2r){\times} SU(k) {\times} U(1)$.  The classical curve
will be corrected by powers of the instanton factor $\Lambda^{2k}$
giving the general form
		\eqn\Espgen{
y^2=x{\til P}^2 + \Lambda^{2k}\til Q(x,\til\phi,\til m)
+\Lambda^{4k}\til R(x,\til\phi,\til m) + \CO(\Lambda^{6k})
		}
where $\til Q$ and $\til R$ are functions whose form is to be
determined.  Break to $Sp(2r){\times} SU(k){\times} U(1)$ by setting
$\til\phi_a = \phi_a$ for $a{=}1,\ldots,r$, and $\til\phi_a =
\phi'_a{+}M$ for $a= r{+}1,\ldots,r{+}k$ with $\sum\phi'_a {=} 0$. We
also set $\til m_j {=} m_j$ to ensure that the $2r{+}2$ quarks remain
light in the $Sp(2r)$ factor.  In the limit $M\gg (\phi_a, \phi'_a,
m_j)$ the three factors decouple.  The semiclassically scale invariant
$Sp(2r)$ factor will have finite coupling $\tau$ if we send $\Lambda
{\to} \infty$ keeping $g\equiv (\Lambda/M)^{2k}$ constant.  At weak
coupling, a one-loop renormalization group matching implies that $g(q)
\propto q$, which can be corrected by higher powers of $q$
nonperturbatively.

In this limit the curve should factorize in a way corresponding to the
decoupling low--energy sectors.  The required factorization implies
that the coefficients of positive powers of $M$ must vanish, so the
${\cal O}(\Lambda^{6k})$ terms in \Espgen\ must vanish, and $\til
R=R(x,m_i)$ cannot depend on $\til\phi_a$. Also, since $\til Q$ is a
symmetric function of the $\til\phi_a^2$, the only terms in $\til Q$
which can survive the limit must have the form
		\eqn\EQmatch{
\til Q = \prod_{a=1}^{r+k} \left( Ax + B\sum m_i^2 - \til\phi_a^2
\right) \cdot Q(x,m_i) + ({\rm lower\ order\ in}\ \til\phi^2),
		}
for some undetermined coefficients $A,B$ and function $Q$.  The terms
lower order in $\til\phi^2$ in general also give contributions
surviving the factorization limit, but, fixing $r$ and taking $k$
arbitrarily large, fewer and fewer of these terms survive.  Since the
$Sp(2r)$ curve should be independent of $k$, the only contributions
must come from the leading term shown in \EQmatch.  Taking $x {\ll} M$
and rescaling $y {\to} M^{2k}y$ in the limit, the $Sp(2r)$ curve thus
becomes
		\eqn\Ecurveform{
y^2 = x P^2 + g P'\cdot Q + g^2 R, \qquad P' \equiv \prod_{a=1}^r
\left( Ax + B \sum m_i^2 - \phi_a^2 \right),
		}
and $Q(x,m_i)$, $R(x,m_i)$ are polynomials invariant under the Weyl
group of the flavor symmetry $SO(4r{+}4)$ of degrees $r{+}1$ and
$2r{+}1$ in $x$, respectively.

Consider now the $Sp(2)$ case ($r{=}1$) with no flavors.  This is the
limit in which we take the four masses $m_i {=} M {\to} \infty$ and the
coupling $\tau \to +i\infty$ such that $q M^4 = \Lambda^4$ is
constant.  The general form for $Q$ in this case is $Q = \alpha x^2 {+}
\beta M^2 x {+} \gamma M^4$ with some unknown constants
$\alpha,\beta,\gamma$.  Then from \Ecurveform\ the $R$ term drops out
in the limit and the $Sp(2)$ no-flavor curve is
		\eqn\Espnoflav{
y^2 = x(x+u)^2 + \gamma (Ax+u) \Lambda^4 + \beta B x \Lambda^4 + \gamma
B \Lambda^4 M^2 ,
		}
where we have defined $u {=} {-}\phi_1^2$, the $Sp(2) {\simeq} SU(2)$
gauge--invariant coordinate on the Coulomb branch.  For this limit to be
consistent, the last term must vanish (since $M {\to}\infty$), so
either $\gamma{=}0$ or $B{=}0$.  Comparing \Espnoflav\ to the $SU(2)$
no flavor curve $y^2 = \til x^2(\til x-u) + \Lambda^4 \til x$ found in
\SWII, we see that they are equivalent only if $B{=}0$, $A{=}1$, and
we shift $x {=} \til x {-} u$.  Thus we learn that $P'=P$ in \Ecurveform.

Consider next the breaking of $Sp(2r)$ with $2r{+}2$ flavors to
$Sp(2r{-}2)$ with $2r$ flavors.  This is achieved by taking $\phi_r ,
m_{2r+1}, m_{2r+2} \sim M {\to} \infty$ and keeping the coupling, and
therefore $g(q)$, finite.  For this limit of \Ecurveform\ not to be
singular, we must have $Q(x,m_i)= C x^{r+1} {+}Dx^r\sum m_i^2 {+}E
\prod m_i$, since all other $SO(2r{+}2)$ invariants would contribute
higher powers of $M$ than $M^2$.  But for $Q$ to preserve its form under this
reduction, we must have $C{=}D{=}0$.  Absorbing the constant $E$ into
our definition of $g$, we have found, so far, that the $Sp(2r)$ curve
with $2r{+}2$ flavors has the form
		\eqn\Ecurveformii{
y^2 = x P^2 + 2 g P\cdot Q + g^2 R, \qquad P \equiv \prod_{a=1}^r
\left( x - \phi_a^2 \right), \qquad Q \equiv \prod_{i=1}^{2r+2} m_i.
		}

To further constrain the curve we construct the one-form $\lambda$.  A
basis of holomorphic one-forms on our hyperelliptic curve are
$\omega_\ell = x^{r-\ell}dx/y$ for $\ell {=} 1,\ldots,r$.  {}From the
definition of the $s_\ell$ \geninvsp, it follows that $\p P/\p s_\ell =
x^{r-\ell}$ and $\p y/\p s_\ell = (1/y) (x^{r-\ell+1}P {+}
x^{r-\ell}Q)$, so it is straightforward to integrate the differential
equation \difeq\ to find the solution, up to a total derivative,
		\eqn\Esplambda{
\lambda=a{dx\over 2\sqrt x}\log\left( {xP + gQ + \sqrt x y \over xP +
gQ - \sqrt x y} \right),
		}
which has logarithmic singularities when
		\eqn\Esingconda{
(xP + gQ + \sqrt x y)\cdot(xP + gQ - \sqrt x y) = g^2(Q^2 - xR)= 0 .
		}
These logs can be converted into poles by adding the total derivative
$d[a\sqrt{x}\log((xP {+} gQ - \sqrt x y)/(xP {+} gQ {+} \sqrt x y))]$
to $\lambda$.  Denoting by $\epsilon^\pm_i$ the roots of the two
factors in \Esingconda, the resulting form has poles $\pm a \sqrt
{\epsilon^\pm_i} dx / (x - \epsilon^\pm_i)$.  For the residues to equal
the masses, we must have $\epsilon^\pm_i \propto m_i^2$.  The flavor
symmetry then implies $Q^2 - xR = \prod (x - m_i^2)$,
where we have rescaled the masses and $x$ to fix the coefficients; this
in turn implies $a {=} 1/2\pi i$.  Putting this all together gives
the curve and one-form \Espcurve.

\subsec{$\tau$-Dependence and S-Duality}

It still remains to determine the coupling constant dependence of the
coefficient $g(q)$.  In principle $g$ could depend on $r$ as well as
on $q$.  We first determine its $r$-dependence by induction in $r$, then
determine the $q$-dependence by matching to the $r{=}1$ case.

The induction proceeds by considering the breaking of $Sp(2r)$ with
$N_f=2r{+}2$ down to $Sp(2r{-}2) {\times} U(1)$ with $2r$ light
hypermultiplets transforming as $({\bf 2r{-}2},0)$. Set
                \eqn\induc{\eqalign{\openup 2\jot
        \phi_a &= \left\{ \matrix{
        \phi'_a\hfill &\qquad\qquad\ \  a=1,\ldots,r{-}1,~\hfill\cr
        M\hfill &\qquad\qquad\ \  a=r,\hfill\cr} \right. \cr
        m_j &= \left\{ \matrix{
        h(q)\, m'_j \hfill&\qquad j=1,\ldots,2r,\hfill\cr
        k(q)\, M \hfill&\qquad j=2r{+}1,\ 2r{+}2,\hfill\cr} \right.
		}}
where $h(q),k(q) = 1 {+} \CO(q)$.  Then the limit $M {\to} \infty$ keeping
$\phi'_a$ and $m'_j$ fixed achieves the desired breaking.  The matching
conditions for the $\phi_a$ in \induc\ define the breaking we are
considering.  The function $h(q)$ in the matching for the masses can be
absorbed in a redefinition of the masses, so can be defined to be $h(q)
{=} 1$.  There is a single mass renormalization $h(q)$
for all the light masses since we are respecting the low--energy global
flavor symmetry which is a simple group.  We are free to choose the
function $k(q)$ as well, as it defines the matching between the
high- and low--energy scale theories.  The simplest choice is $k(q){=}1$.
One then finds that
\Espcurve\ reduces to a curve of the same form with $r\to r{-}1$ and
$g_{r-1}(q_{r-1}) = g_r(q_r)$.  The one-loop  renormalization group
matching condition that $g_r(q_r) \propto q_r$ independent of $r$
implies that the bare couplings satisfy $\tau_{r-1} {=} \tau_r$ at weak
coupling.  Nonperturbatively this relation can be modified by positive
powers of $q_r$ \MNcorr: $q_{r-1} {=} q_r \ell(q_r)$ with $\ell(q) =
1{+}\CO(q)$.  The choice of the function $\ell(q)$ is arbitrary; one
can view it as a renormalization prescription defining what is meant by
the coupling nonperturbatively.  Our prescription will be to choose
$\ell(q) {=} 1$, in other words we choose $\tau_{r-1} {=} \tau_r$
nonperturbatively.

It is clear that the above renormalization prescription is consistent;
however, many other consistent possibilities exist.  For instance,
matching with $k{\neq}1$ (and $\ell{=}1$) gives $g_{r-1} {=} k^2 g_r$.
If the function $k$ is chosen to be singular enough, $g_{r-1}$ can have
different modular properties than $g_r$.  (In particular, $k(q)$ will
need to have infinitely many poles in the $|q|{<}1$ disk, which
accumulate on the boundary.)  Which coupling dependence $g(q)$ of our
curves is the right one?  One cannot answer this question in the
present framework since there we have no independent definition of what
the coupling $\tau$ means (away from $\tau = +i\infty$).  If one were
directly computing the effective theory from a specific ({\it e.g.}
lattice, string theory) regularization of the theory at high energies,
then there would be a correct answer;  however, this answer could
depend on the regulator.  With the less direct methods
we have at our disposal at present, we will be content to take the
above, simplest, matching condition to determine the strong--coupling
and modular behavior of our scale invariant solutions.

We determine the unknown function $g(q)$ by matching to Seiberg and
Witten's $SU(2) {\simeq} Sp(2)$ solution.  Take the 4-flavor curve as
given in Eq.~(16.38) of \SWII, and make the following redefinitions,
using their notation:
		\eqn\Eswredef{\eqalign{
	y^2 &\rightarrow {c_1^2\over c_1^2{-}c_2^2} y^2 ,\qquad
	x \rightarrow x+ {c_1^2\over c_1^2{-}c_2^2} u ,\cr
	u\ &\rightarrow {c_1\over c_1^2{-}c_2^2} u ,\qquad
	m_i \rightarrow {1\over \sqrt{c_1^2{-}c_2^2}} m_i .\cr
		}}
Then their curve becomes precisely the $r{=}1$ curve in \Espcurve\ with
$g = (c_2/c_1) = \vartheta_2^4/(\vartheta_3^4 {+} \vartheta_4^4)$,
in terms of the usual Jacobi theta functions \thetafnc.

It follows from \thetaST\ that $g$ is invariant under $T^2$ and
$ST^2S$, while $T{:}\ g \to -g$.  The curve is invariant under this
sign change if, at the same time, the sign of a {\it single} mass is
changed.  Note that this sign change is not part of the nonanomalous
$SO(4r{+}4)$ flavor symmetry (whose Weyl group contains only pairwise
sign flips of the masses), but instead is the $\bZ_2$
outer automorphism of the group.  $T$ and $ST^2S$ generate a duality
group $\Gamma_0 \subset PSL(2,\bZ)$ which can be characterized as the
set of $SL(2,\bZ)$ matrices whose lower off--diagonal element is even.
This should be contrasted with the $Sp(2)$ case \SWII, where the
duality group is all of $PSL(2,\bZ)$, and is mixed with the $S_3$
outer automorphisms of the $SO(8)$ flavor group.

\subsec{Checks}

In the course of deriving the form of the $Sp(2r)$ curve above, we
checked that the adjoint breaking $Sp(2r){\to}Sp(2r{-}2){\times}U(1)$
was consistently reproduced by our solution.  We now check that the
other adjoint breaking $Sp(2r){\to}SU(r){\times}U(1)$ is also reproduced.
The semiclassical breaking of $Sp(2r)$ with $N_f=2r{+}2$ down to 
$SU(r)\times U(1)$ with $2r$ light hypermultiplets transforming as
$({\bf r},0)$ is achieved by tuning
                \eqn\spchk{\eqalign{
        \phi_a &= M {+} \til\phi_a \qquad\qquad\qquad\qquad 
	\sum_{a=1}^r\til\phi_a=0,\cr
        m_j &= \left\{ \matrix{M {+} \til m_j {+} 2h(q)\til\mu \hfill&\qquad 
	\til\mu \equiv {1\over2r}\sum_{j=1}^{2r}\til m_j,\hfill\cr
        0 \hfill&\qquad j=2r{+}1,\ 2r{+}2,\hfill\cr} \right.
		}}
in the limit $M {\to} \infty$ keeping $\til\phi_a$ and $\til m_j$ fixed,
and where $h(q) {\sim} \CO(q)$.  The relative renormalization $h(q)$
of the flavor--singlet mass reflects the fact that the global flavor
symmetry of the $SU(r)$ theory is not a simple group.
Substituting \spchk\ into \Espcurve, shifting $x \to M^2{+}2M\til x$, 
$y \to M^{r+1}\til y$, expanding around $|\til x|{\ll}|M|$, and tuning
$h(q)$ appropriately, we indeed recover the $SU(r)$ curve \Esucurve.
The other maximal adjoint breakings, $Sp(2r) \to Sp(2r{-}2k){\times}
SU(k){\times} U(1)$, are equally easy to check.

The $Sp(2)$ curve should also be equivalent to the $SU(2)$ curve
\Esucurve.  If we write the latter curve as
		\eqn\sutwocrv{
y^2 = (x^2-u)^2 +4h(h{+}1)\prod_{j=1}^4\left(x-m_j-\half h (
\textstyle{\sum_k m_k})\right),
		}
where $u=-\phi_1\phi_2=\phi_1^2$, and the $Sp(2)$ curve \Espcurve\ as
		\eqn\sptwocrv{
xy^2 = \left(x(x-\til u)^2+g\textstyle{\prod_j\til m_j}\right)^2
-g^2 \prod_{j=1}^4(x-\til m_j^2),
		}
with $\til u = \phi_1^2$, then the discriminants of the two curves
are the same if we relate the parameters by $\til u = \beta^2[4u -
h(h{+}1)(\sum_j m_j)^2+h\sum_j m_j^2]$, $\til m_j = 2\beta (1 {+} 
{3\over2} h)^{1/2} m_j$ with $\beta = (1{+}h) (1{+}{3\over2}h) (1{+}2h)$.
This reproduces the expected weak--coupling matching as $h{\to}0$.
The equality of the discriminants for these two tori imply that they
are related by an $SL(2,\bC)$ transformation of $x$, and incidentally
shows the equivalence of the $SU(2)$ solution \Esucurve\ with the 
results of \SWII.

Another check on the validity of our solution is that it correctly
reproduces the positions and monodromies of singularities at weak coupling.  
We will check two classes of such singularities:  the gauge singularities 
which correspond classically to the restoration of a nonabelian gauge 
symmetry, and the quark singularities which correspond to hypermultiplets 
becoming massless.  

For $Sp(2r)$ the gauge singularities occur whenever $\phi_a^2 {=}
\phi_b^2$ or $\phi_a{=}0$.  Because the beta function vanishes, the
semiclassical monodromies around the gauge singularities are actually
the classical monodromies given by elements of the Weyl group of
$Sp(2r)$, which act by permuting the $\phi_a$'s or flipping their
signs.  The breakings
\induc\ and \spchk\ imply that all the $Sp(2r{-}2)$ and $SU(r)$
singularities and associated monodromies are reproduced by the $Sp(2r)$
curve, allowing us to check the gauge monodromies by induction in $r$.
We need only compute for $Sp(2r)$ a generating
monodromy not contained in the Weyl
group of $Sp(2r{-}2)$.  A convenient choice is a Coxeter element of
the Weyl group \refs{\KLTYii,\UDBS} corresponding to a cyclic
permutation of the $\phi_a$ and a sign change of one element, which
gives the monodromy ${\bf S} = {\t P^{-1}\ 0\choose\ 0\ \ P}$,
where $P$ is the $r{\times}r$ matrix representation of the Coxeter element
({\it i.e.}\ its action on the $\phi_a$'s).

For weak coupling, $|q|{\ll}1$, and vevs much larger than the bare
masses $\phi_a{\gg}m_i$, the curve is approximately $y^2 = x \prod
(x{-}\phi^2_a)^2 - q^2 x^{2r+1}$.  Degenerations where two branch
points collide occur whenever $\phi^2_a {=} \phi^2_b$ or $\phi^2_a {=}
0$, up to corrections of order $q$, corresponding to the semiclassical
positions of the gauge singularities.  The special monodromies can be
conveniently measured by traversing a large circle in the $s_r$ complex
plane, fixing the other $s_\ell {=} 0$, where the curve factorizes as $y^2
= x[(1{-}q)x^r {+} s_r]\cdot [(1{+}q)x^r {+} s_r]$.  The branch points
are arranged in $r$ pairs with a pair near each $r$th root of unity times
$s_r^{1/r}$.  As $s_r {\to} e^{2\pi i} s_r$, these pairs rotate into
one another in a counterclockwise sense.  In addition there is one
branch cut extending from the origin to infinity.  Choose cuts and a
basis for the cycles as in the corresponding argument for the $SU(r)$
curve \APS.  As $s_r {\to} e^{2\pi i}s_r$ the cycles are cyclically
permuted, and a pair of them pick up a minus sign as they pass through
the extra cut extending from the origin, thus giving the the classical
monodromy predicted above.

Classically, quark singularities occur whenever $\phi_a {=} \pm m_i/
\sqrt2$, corresponding to the $q^i_a$, $\tq^a_i$ hypermultiplets becoming
massless.  In the effective theory, the massless quark can be taken to
have electric charge one with respect to a single $U(1)$ factor and to
carry quark numbers $n^j = \delta^j_1$.  The semiclassical monodromy
around the quark singularity can be read off from \dyonmon.

Consider the curve near a classical quark singularity,
say $\phi_1 {\sim} m_1/\sqrt2$.  At weak coupling and for $x {\sim}
\phi^2_1$ the curve is approximately $y^2 = [C_1(x {-} \phi^2_1) {-}
qC_2]^2 - q^2 C_3 (x {-} m_1^2)$, where the $C_i$ are slowly--varying
functions of $x$ and $s_\ell$. This has a double zero at $x = \half
m^2_1 {+} \half q^2 (C_2/C_1^2)$ for $\phi^2_1 = \half m^2_1 {-}
q(C_2/C_1) {-} \half q^2 (C_3/C_1^2)$, which is indeed near the
classical quark singularity for small $q$.  Define the period $a_1$ by
a contour enclosing the pole at $m_1^2$ (recall that changing which
poles are enclosed by a given contour corresponds to a physically
unobservable redefinition of the quark number charges).  One then finds
that as $\phi_1$ winds around the singular point the two branch points
are interchanged.  The monodromy which follows from this is nontrivial
only in a $2{\times}2$ block of \dyonmon, for which we find $\bS =
{1\ 1\choose0\ 1}$ and $\bT={1\choose0}$, in agreement with the
semiclassical prediction.

\newsec{SO(2r+1)}

The arguments in this case are essentially the same as in the
$Sp(2r)$ case, so we will run through them more quickly.  Also, there
is some simplification compared to the $Sp(2r)$ case due to the fact
that the flavor symmetry in the $SO(n)$ case does not admit the
``extra'' invariant $\prod m_j$.

\subsec{Symmetries}

$SO(2r{+}1)$ has rank $r$ and dimension $r(2r{+}1)$, its adjoint
representation has index $4r{-}2$, and the $2r{+}1$-dimensional vector
representation is real and has index $2$.  The beta function
for the theory with $N_f$ vector hypermultiplets is then $i\pi\beta =
2N_f {-} 4r {+} 2$, and the instanton factor is $\Lambda^{4r-2-2N_f}$.

On the Coulomb branch, the adjoint chiral superfield $\Phi$ 
expectation value can be skew--diagonalized as
		\eqn\Ephisp{
	\vev\phi = \pmatrix{
	0 &\phi_1 & & & & \cr
	-\phi_1 &0& & & & \cr
	& &\ddots   & & & \cr
	& & &0 &\phi_r  & \cr
	& & &-\phi_r  &0& \cr
	& & &         & &0\cr} .
		}
The gauge invariant combinations of the $\phi_a$'s are all the
symmetric polynomials in $\phi_a^2$ up to degree $2r$.  The generating
polynomial for these invariants is $\prod_a^r (x -\phi_a^2)\equiv
\sum_{\ell=0}^r s_\ell x^{r-\ell}$.  The general (maximal) adjoint
breaking is $SO(2r{+}1) {\to} SO(2k{+}1) {\times} SU(r{-}k) {\times}
U(1)$, which occurs via an expectation $\vev\phi = (M, \ldots, M, 0,
\ldots, 0)$ with $r{-}k$ $M$'s.  The vector decomposes as $\bf 2r{+}1 =
( r{-}k, 1) {\oplus} (\overline{r{-}k}, 1) {\oplus} (1,2k{+}1)$ under
$SU(r{-}k) {\times} SO(2k{+}1)$.

The flavor symmetry is $Sp(2N_f)$ and the $\CN{=}2$ invariant masses
transform in the adjoint representation, and can be diagonalized to
$\pm m_i$, $i {=} 1,\ldots,N_f$.  The $Sp(2N_f)$ invariants are all the
symmetric polynomials in the $m_i^2$ up to degree $2N_f$.  To see the
$Sp(2N_f)$ flavor symmetry, define $X=(Q,\til Q)$, so ${\cal W} = X_a
{\cdot} J {\cdot} \t X_b \,\Phi^{ab}$ where $J^s_r = {\ 0\ 1\choose
-1\ 0} {\otimes} \bo_{N_f}$, the symplectic metric, since the adjoint
$\Phi^{ab}$ is antisymmetric.  So, the global symmetry is $U(2N_f) \cap
Sp(2N_f,\bC) \equiv U\!Sp(2N_f)$, the unitary symplectic group.

\subsec{Curve and One-Form}

As in the $Sp(2r)$ case, we impose that the curve be hyperelliptic of
the form \Ecrvform.  The same argument gives the ``classical'' piece as
\Eclguessii.  In the scale invariant case with $2r{-}1$ masses there
will be an overall factor of $x^\epsilon$ in the curve.  
$\epsilon{=}2$ would make the curve singular everywhere, so we must
take $\epsilon{=}1$.  The ``quantum'' piece is determined by
considering the most general curve for $SO(4r{+}1)$ with $2r{-}1$
flavors and breaking to $SU(r) {\times} SO(2r{+}1) {\times} U(1)$ at a
large scale $M$, by letting $\phi_a {\to} M$ for $a{=}1,\ldots,r$.  In
the limit $M {\gg} (\phi_a, m_j)$ the three factors decouple. To obtain
the $SO(2r{+}1)$ factor with $2r{-}1$ flavors at finite coupling
$\tau$, we should send $\Lambda {\to} \infty$ such that
$(\Lambda/M)^{4r} {\equiv} f(q) {\sim} \CO(q)$, by a one-loop
renormalization group matching.  Taking the limit $M{\to}\infty$, the
classical piece factorizes into a piece with $2r{+}1$ zeros near
$x{=}0$ (relative to the scale $M$) and another piece with zeros all of
order $M$.  The whole curve should factorize in this way to correspond
to the decoupling low--energy sectors.  This implies that the
scale invariant curve must have the form $y^2 = xP^2 {+} 4f Q$ where $Q
= Q(x, m^2_j)$ is symmetric in the $m^2_j$ and homogeneous of degree
$2r$.  $Q$ is fixed by integrating \difeq\ to find $\lambda \propto
(dx/\sqrt x) \log [(xP {-} \sqrt xy)/(xP {+} \sqrt xy)]$. The
logarithmic singularities at $x^2=\epsilon_j$, the zeros of $Q$, are
converted into poles by integrating by parts.  The residues of
$\lambda$ are linear in the quark masses if $\epsilon_j \propto m_j^2$,
and the only flavor--symmetric $Q$ with this property is $Q=x^2 \prod_j
(x {-} m^2_j)$.  Putting this all together gives the $SO(2r{+}1)$ curve
and one-form \Esooddcurve.

\subsec{$\tau$-Dependence and S-Duality}

We still need to determine $f(q)$ in \Esooddcurve.  In principle, $f$
could depend on $r$ as well as $q$.  We determine its $r$-dependence 
by considering the breaking of $SO(2r{+}1)$ with
$N_f=2r{-}1$ down to $SO(2r{-}1){\times} U(1)$ with $2r{-}3$ light
hypermultiplets transforming as $({\bf 2r{-}1},0)$.  To this end, set
the parameters as in \induc.  In the limit $M{\to}\infty$ one finds
that the $SO(2r{+}1)$ curve reduces to a curve of the same form with $r
\to r{-}1$ and $f_{r-1} {=} f_r$.  The one-loop  renormalization
group matching condition that $f(q) \propto q$ independent of $r$
implies that the bare couplings satisfy $\tau_r = \tau_{r-1}$ at weak
coupling.  We choose this as our renormalization prescription at strong
coupling as well.

We determine the unknown function $f(q)$ by matching to the solution
\SWII\ of the $SU(2)$ theory with a massless adjoint hypermultiplet.
It will be easiest to match to the $SU(2)$ curve in the form \Esucurve.  
According to \SWII, the $SU(2)$ solution with a single adjoint
hypermultiplet
of mass $\tm$ is given by the 4-fundamental-flavor solution with masses
$(\tm,\tm,0,0)$.  With these masses, \Esucurve\ becomes
		\eqn\suadj{
y^2 = (x^2-u)^2 + 4h(h{+}1)(x-(h{+}1)\tm)^2(x-h\tm)^2,
		}
where we have defined the gauge--invariant coordinate on the
Coulomb branch as $u=-\phi_1\phi_2=\phi_1^2$.
The discriminant of \suadj\ is
		\eqn\discsuadj{
\Delta_{su} = 4096h^2(h{+}1)^2(2h{+}1)^2(u-h^2\tm^2)^2(u-h(h{+}1)\tm^2)^2
(u-(h{+}1)^2\tm^2)^2 .
		}
Our $SO(3)$ curve, on the other hand, is
		\eqn\sothree{
y^2 = x(x-v)^2 + 4fx^2(x-m^2),
		}
where $v=\phi_1^2$ is the gauge--invariant coordinate, and its
discriminant is
		\eqn\discsothree{
\Delta_{so} = -16fv^4(v^2-m^2v-fm^4).
		}
The most general relation between the coordinates and
masses allowed by dimensional considerations and agreeing
with the weak--coupling limit is
		\eqn\sosumatch{
v=A(q) u + B(q) \tm^2, \qquad m^2 = C(q) \tm^2, 
		}
where $A(q), C(q) = 1{+}\CO(q)$ and $B(q) \sim \CO(q)$.

Now, if the two tori \suadj\ and \sothree\ are equivalent by an
$SL(2,\bC)$ transformation of $x$, then their discriminants should
be equal for some $f(q)$ after a suitable change of variables
\sosumatch.  However, it is clear that this is impossible, since
\discsuadj\ has three double zeros in $u$, while \discsothree\
has one quartic and two single zeros in $v$.
Nevertheless, the two curves are equivalent, being related by an
isogeny:  for fixed values of the parameters, each is a double
cover of the other.  (An example of isogenous descriptions of the
same physics appeared in \SWII.)  More explicitly,
the double cover of the $SO(3)$ torus \sothree\ is given by
		\eqn\isosothree{
y^2 = (z^2-v)^2 + 4fz^2(z^2-m^2),
		}
which is obtained from \sothree\ by dropping the overall factor of
$x$ and then replacing $x\to z^2$.  The fact that \sothree\
is cubic in $x$ and has an overall factor of $x$ implies that two
of its four branch points are at $\infty$ and $0$ on the $x$-plane,
independent of the values of the parameters.  (By an $SL(2,\bC)$
transformation any torus can be brought to this form.)  Moving to
a double cover of the $x$-plane by $x=z^2$ then effectively removes the
branch points at $x=0,\infty$, leaving the isogenous curve \isosothree.
Its discriminant is
		\eqn\discisosothree{
\Delta_{\til{so}} = 4096f^2(1{+}4f)v^2(v^2-m^2v-fm^4)^2.
		}
This indeed becomes \discsuadj\ under a change of variables
\sosumatch\ with $A = C = 1$, $B = h(h{+}1)$, and $f = h(h{+}1) =
\vartheta_2^4 \vartheta_4^4/ (\vartheta_2^4 {-} \vartheta_4^4)^2$.

The modular transformation properties of the theta functions \thetaST\ 
imply that $f$, and thus the $SO(2r{+}1)$ curve and
one-form, are invariant under $T^2$ and $S$, which generate a duality
group $\Gamma^0(2) \subset PSL(2,\bZ)$ characterized as the set of
$SL(2,\bZ)$ matrices with even upper off--diagonal element.
For $r{=}1$, the curve is in fact invariant under all of $PSL(2,\bZ)$,
though this is not manifest in the form given in \Esooddcurve.  One
way of seeing this is to note that (by an argument similar to the
one given above) the $SO(3)$ curve is isogenous to the $Sp(2)$
curve with masses $(m,m,0,0)$, and this can be rewritten in a 
manifestly $SL(2,\bZ)$-invariant form using the change of variables
\Eswredef.

\subsec{Checks}

It is easy to check that the curve \Esooddcurve\ reproduces the
maximal adjoint breaking patterns $SO(2r{+}1) \to SO(2r{-}2k{+}1)
{\times} SU(k) {\times} U(1)$ in a manner similar to the analogous
check for the $Sp(2r)$ curves in section 3.4.  The positions and
monodromies of gauge and quark singularities also match with
perturbation theory at weak coupling.  Here the gauge singularity
monodromy is precisely the same as in the $Sp(2r)$ case, since both
curves have the same limit at weak coupling and small bare quark
masses, and because their Weyl groups are the same.  Checking the
quark singularities also involves a calculation very similar to
(though slightly simpler than) the $Sp(2r)$ case, which we will not
repeat.

Finally, in the limit that we take all the bare quark masses large
(and $q{\to}0$ appropriately), our curve should describe pure $SO(2r{+}1)$
Yang--Mills theory.  The curve is, in this limit,
		\eqn\sooddym{
y^2 = x\prod_{a=1}^r(x-\phi_a^2)^2 + x^2 \Lambda^{4r-2}.
		}
A seemingly different $SO(2r{+}1)$ Yang--Mills curve,
		\eqn\sooddymDS{
y^2 = \prod_{a=1}^r(z^2-\phi_a^2)^2 + z^2 \Lambda^{4r-2},
		}
was proposed in \UDBS.  In fact, the two curves are equivalent, the
second being a double cover of the first.  Indeed, one can transform
the first curve into the second by the same ``isogeny'' change of
variables as we used to go from \sothree\ to \isosothree.  In general,
this transformation takes us from a genus-$r$ curve to a genus
$2r{-}1$ curve with a $\bZ_2$ symmetry.  In order for the
higher--genus curve to reproduce the periods (and hence the physics)
of the lower--genus curve, we must divide by the $\bZ_2$ symmetry just
as was done in \UDBS.

\newsec{SO(2r)}

The argument and results for $SO(2r)$ closely parallel those of
$SO(2r{+}1)$.  Since $SO(2r)$ can not be obtained from $SO(2r{+}1)$ by
adjoint symmetry breaking, we need to give a new induction and
matching argument. In section 6, we will see how to obtain the
$SO(2r)$ curve directly from $SO(2r{+}1)$ by giving an expectation
value to a squark in the vector representation.

The adjoint representation has dimension $r(2r-1)$ and index $4r-4$;
the vector representation has dimension $2r$ and index 2. Thus the
${\cal N}{=}2$ beta function is $i\pi\beta=2N_f-4r+4$ and the
instanton factor is $\Lambda^{4r-4-2N_f}$.  As we did for
$SO(2r{+}1)$, let $\phi_1,\ldots,\phi_r$ be the skew--diagonal entries
of the $2r{\times} 2r$ matrix $\vev\phi$.  The Weyl group is generated
by permutations and by simultaneous sign changes of pairs of the
$\phi_a$, so the symmetric polynomials $s_\ell$ of the $\phi_a^2$,
generated by $\sum_\ell s_\ell x^{r-\ell} = \prod_a(x {-} \phi_a^2)$,
are gauge-invariant.  In addition to the $s_\ell$, there is one
``extra'' Weyl invariant $t=\phi_1\phi_2\cdots\phi_r$, which might be
expected to appear in the curve.  However, $SO(2r)$ also possesses an
outer automorphism corresponding to reflection of the Dynkin diagram
about its principal axis, which interchanges the two spinor roots and
takes $t\to -t$. This additional global symmetry means that our curve
can be taken to depend only on $s_r = t^2$.  As was the case for
$SO(2r{+}1)$, the flavor symmetry is $Sp(2N_f)$; hence the
flavor--invariant mass combinations are again the symmetric
polynomials in the masses, up to degree $2N_f$.

By following the argument of the previous section, we can deduce that
the scale invariant curve takes the form $y^2=xP^2-g^2Q$ with $P$ as
before and $Q=Q(x,m_j^2)$ a symmetric function of the $m_j$ of degree
$2r{+}1$ in $x$. The differential form $\lambda$ is the same as in the
$SO(2r{+}1)$ case, implying that $Q$ has zeroes at $x=m_j^2$ and
therefore that $Q=x^3\prod_{j=1}^{2r-2}(x{-}m_j^2)$. We thus obtain a 
curve of the form \Esoevencurve.  One unexpected feature of this
solution is that the ``classical'' piece of the curve $xP = x\prod (x
{-} \phi_a)$ has singularities whenever any one $\phi_a{=}0$, in
addition to singularities whenever $\phi_a {=} \phi_b$.  The latter
corresponds to an enhanced gauge group classically, but there is no
such enhanced symmetry when $\phi_a{=}0$.  Therefore, for this curve to
be correct there must be no monodromy ({\it i.e.}, only a trivial
monodromy) around such singularities.  In terms of 
gauge invariant parameters, $\phi_a {=} 0$ means that $t{=}0$; 
for $t\sim 0$ the curve near $x \sim 0$
is approximately $y^2 = x(A_1x{-}t^2)(A_2x{-}t^2)$ with nonzero
constants $A_i$.  As $t \to e^{2\pi i}t$, the branch points at
$x=t^2/A_i$ wind twice around the origin, and a simple contour
deformation argument shows that the resulting monodromy is, in fact,
trivial.

The same induction argument as for $SO(2r{+}1)$ implies that we can
take $f_r(q) {=} f_{r-1}(q)$; to find $f(q)$, we study the  breaking of
$SO(2r)$ to $SU(k)$ and match to eq. \Esucurve.  Consider the breaking
of $SO(2r{+}2k)$ with the critical number of flavors $2r{+}2k{-}2$, to
$SU(k){\times}SO(2r)$ with $2k$ and $2r{+}2$ flavors, respectively.  The
breaking is achieved by taking $\til \phi_a {=} \phi_a$ for $a {=}
1,\ldots,k$ and $\til \phi_b {=} M {+} \phi_b'$ for $b = k{+}1, \ldots,
k{+}r$, with $\sum_{r{+}1}^{r{+}k} \til\phi_b = 0$. In order for the
resulting $SU(k)$ and $SO(2r)$ limits to be critical, we must also
shift the masses $\til m_i {=} m_i$ for $i = 1,\ldots,2r{-}2$ and $\til
m_j = M {+} m_j {+} r(q)\mu$ for $j = 2r{-}1,\ldots, 2r {+} 2k {-} 2$.
The resulting curve is
		\eqn\Esoevenbreak{
y^2=\prod_1^r(x-\phi_a^2)^2\prod_1^k(x-(\phi_b+M)^2)^2
+fx^3\prod_1^{2r-2}(x-m_i^2)\prod_1^{2k}(x-(M+m_j+r(q)\mu)^2)
		}
By expanding near $x\sim 0$, we readily recover an $SO(2r)$ curve of
the same form as the $SO(2r{+}2k)$ curve.  To obtain $SU(k)$, we expand
around $M^2$
using $x =   M^2{+}2M\til x$. In the large-$M$ limit, the overall
$M$-dependence factors out, leaving just
		\eqn\Esoeventosu{
y^2=\prod_1^k( x-\phi_b)^2 + f(q)\prod_1^{2k}(x-m_j-r(q)\mu)
		}
This is exactly the curve \Esucurve, with $f = 4h(h{+}1)$ and
$r(q)=2h(q)$.  We have thus found the complete form of the curve for
$SO(2r)$. This result will be confirmed in the next section.

It is easily checked that the positions and monodromies of the
semiclassical singularities of this curve match perturbation
theory, by an argument similar to that given for $Sp(2r)$.  The only
subtlety which arises is that the Coxeter monodromy is generated by
traversing a large circle in the $t$-plane, which corresponds to
traversing a large circle {\it twice} in the $s_r$-plane.  The
resulting monodromy corresponds to a cyclic permutation of the
$\alpha$-cycles times a sign flip of two of them---precisely the
Coxeter element of the $SO(2r)$ Weyl group which only includes
pair-wise sign flips.  Finally, the $SO(2r)$ Yang--Mills (no flavors)
curve found in \BL\ is simply a double cover of the one derived from
\Esoevencurve\ by sending the bare quark masses to infinity at weak
coupling.

\newsec{Higgs Breaking}

We now perform another check on the curves (1.1)-(1.4), this time
coming from physics on the Higgs branches of these theories.  This
argument depends on a nonrenormalization theorem {\nrt{}, which states
that the prepotential ${\cal F}(A)$ determining the low--energy
effective action can have no dependence on the vev of any
hypermultiplet.  The theorem is proved by considering the form of the
most general low--energy effective action for $\CN{=}2$ vector and
hypermultiplets \dWLVP.  It implies in particular that the low--energy
theory along any Coulomb or mixed Higgs--Coulomb branch can not depend
on the squark vevs. We will use this theorem to extend a solution
valid for large squarks (and weak coupling) to arbitrary squark
expectation values, including those that correspond to strong
coupling.

\subsec{SU(n)}

We first write down the $F$- and $D$-term equations which describe the
classical moduli space of the $SU(n)$ theory.  Denote, as usual, the
hypermultiplet vevs by $q^i_a$ and $\tq^a_i$, and the vector multiplet
vev by the traceless $\phi^a_b$, where $a{=}1,\ldots,n$ is a color
index and $i{=}1,\ldots,N_f$ is the flavor index.  The $F$- and
$D$-terms are then $[\phi,\phi^\dagger]=0$ determining the Coulomb
branch (where $q{=}\tq{=}0$), $q^i_a\tq^b_i = q^i_a (q^\dagger)^b_i
{-} ({\tq}^\dagger)^i_a \tq^b_i \propto \delta^b_a$ determining the
Higgs branch (when $\phi{=}0$), and $q^j_a m^i_j {+} \phi^b_a q^i_b =
m^j_i \tq^a_j {+} \tq^b_i \phi^a_b = 0$ governing the mixed
Higgs--Coulomb branch.  Here $m^i_j$ denotes the quark mass matrix
which is in the adjoint of the $U(N_f)$ flavor group.

We are interested in the simplest nontrivial solution to the Higgs
equations, namely $q^i_a = \vev q \delta^i_{N_f-1} \delta_a^n$, $\tq^a_i
= \vev q \delta_i^{N_f} \delta^a_n$.  Along this direction on the Higgs
branch only two flavors of squark get a vev, breaking the gauge group
to $SO(n{-}1)$.  For the mixed Higgs--Coulomb equations to be satisfied,
the masses of these two squarks must vanish $m_{N_f-1} {=} m_{N_f} {=} 0$.
It is then clear that the mixed Higgs--Coulomb equations admit solutions
with nonzero $\phi^a_b$ satisfying the Coulomb equation and the condition
$\phi^n_a {=} \phi^b_n {=} 0$ for all $a,b$.  This condition simply
reduces the rank of $\phi$ so that it describes the Coulomb branch of
the $SU(n{-}1)$ unbroken by $\vev q$.

Physically, we have shown that there is a classical flat direction along 
which two quarks get a vev $\vev q$, Higgsing $SU(n) {\to} SU(n{-}1)$ and
reducing the number of light flavors from $N_f$ to $N_f{-}2$.  We expect
this picture to be quantum--mechanically accurate only in the limit
$\vev q {\to} \infty$, where the physics on the Higgs branch becomes
arbitrarily weakly coupled.  Thus, in this limit, when two of the
bare masses $m_{N_f-1} {=} m_{N_f} {=} 0$, 
the Coulomb branch for $SU(n{-}1)$ with $N_f{-}2$ flavors 
emanates from the Higgs branch.  By the nonrenormalization theorem
stated above, this $SU(n{-}1)$ Coulomb branch cannot depend on the
value of $\vev q$.  So we are free to take the limit $\vev q {\to} 0$,
which identifies the $SU(n{-}1)$ Coulomb branch as the ``root'' of the
mixed branch where it intersects the $SU(n)$ Coulomb branch (see Fig.~1).  
This intersection is determined from the curve \Esucurve\ for 
$SU(n)$ as the submanifold where the renormalized mass of two quarks is
zero ({\it i.e.} the submanifold of points from which a Higgs branch can
emanate).\fig{Crude sketch of the quantum moduli space of $SU(n)$
showing the intersection of the Coulomb and a mixed Higgs--Coulomb
branch.  A nonrenormalization theorem implies that the low--energy
effective action at this intersection is the same as the effective
action of the $SU(n{-}1)$ theory far out along the mixed branch.}{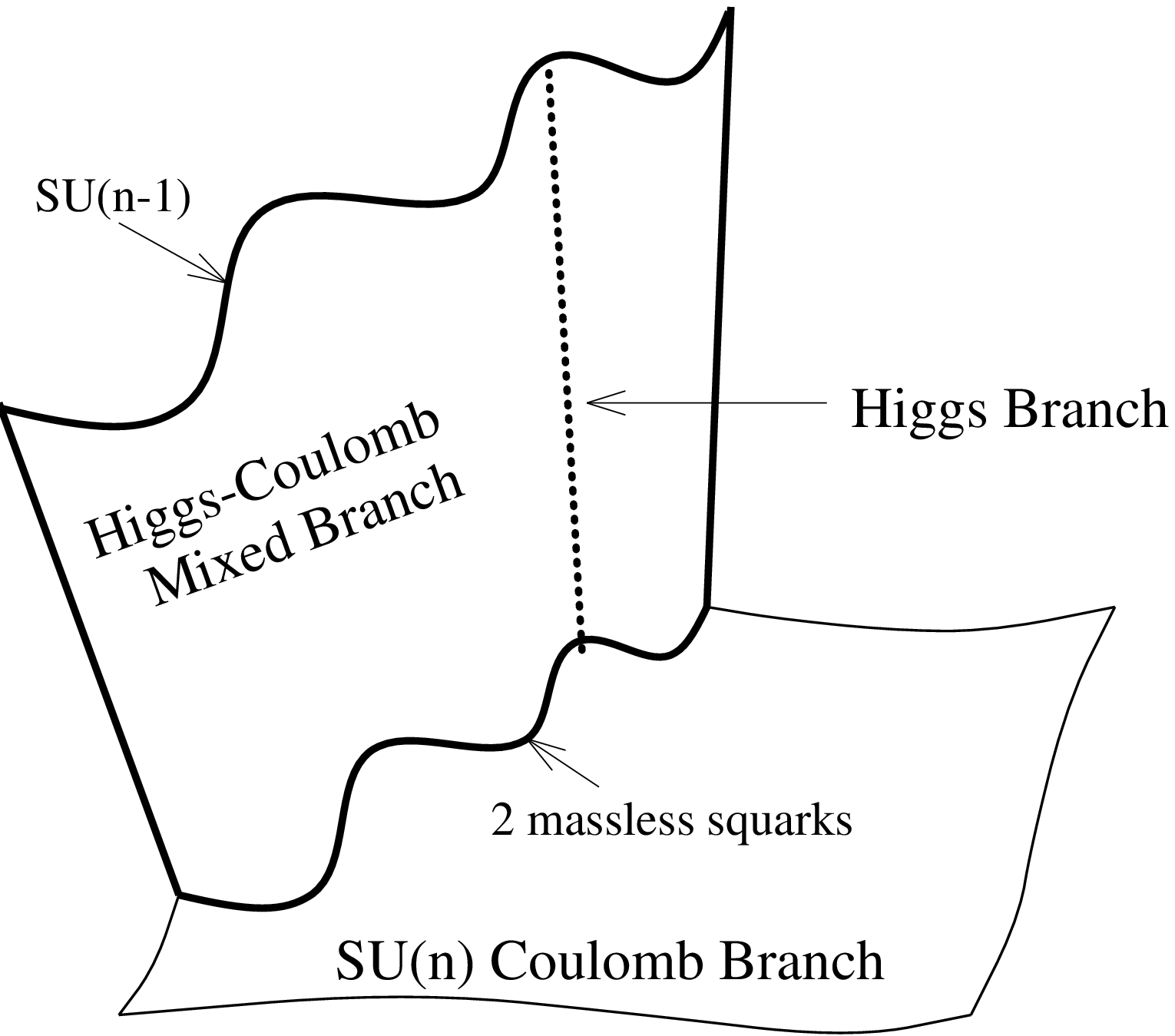}
{8cm}

It is easy to find this submanifold explicitly from our curve.  The
$SU(n)$ curve \Esucurve\ with two bare masses set to zero is
		\eqn\Hsu{
y^2 = \prod_{a=1}^n (x - \phi_a)^2 + 4h(h+1) (x-2h\mu)^2 
\prod_{j=1}^{2n-2} (x - m_j - 2h \mu) ,
		}
where $\sum \phi_a = 0$ and $\mu \equiv (1/2n) \sum m_j$.
There is a quark singularity precisely when $\phi_n = 2h\mu$, in
which case the curve becomes the singular piece $(x {-} 2h\mu)^2$
times the curve
		\eqn\Hsuii{
y^2 = \prod_{a=1}^{n-1} (x - \phi_a)^2 + 4h(h+1) 
\prod_{j=1}^{2n-2} (x - m_j - 2h \mu) ,
		}
which by the above argument we should identify as the $SU(n{-}1)$
curve with $2n{-}2$ flavors.  Redefining $x = \til x - 2h\mu/(n{-}1)$,
$\phi_a = \til\phi_a - 2h\mu/(n{-}1)$, and $\mu = n\til\mu/(n{-}1)$,
we see that \Hsuii\ indeed becomes precisely the $SU(n{-}1)$ curve.

\subsec{SO(n)}

The $SO(n)$ case is simpler.  Again assembling the squark vevs into
$2N_f$-component vectors $X^a_i$, the Higgs--branch equations are $\bar
X^{[a}_iX^{b]}_i = X^a_i J^{ij} X^b_j =0$ where $J$ is the symplectic
metric.  The moduli space has flat directions for $X^a_i$ having a
single non-zero entry with the associated bare mass zero.  Such a vev
parameterizes a mixed Higgs--Coulomb branch, along which $SO(n)$ is
broken to $SO(n{-}1)$ and the Higgs mechanism lifts one of the
flavors.  By the nonrenormalization theorem, we can identify the
$SO(n{-}1)$ curve from the $SO(n)$ curve with one bare mass set to
zero, by looking at the intersection of the $SO(n)$ Coulomb branch
with the mixed branch.

When breaking $SO(2r) \to SO(2r{-}1)$, we need to tune $\phi_r {=}0$
in \Esoevencurve\ to find this intersection (where a single quark is
massless).  Factoring out the $x^2$ singularity indeed gives the
$SO(2r{-}1)$ curve \Esooddcurve.  When breaking $SO(2r{+}1) \to
SO(2r)$ we do not need to tune the $\phi_a$'s at all since the two
groups have the same rank.  Thus we learn that the whole $SO(2r{+}1)$
Coulomb branch with one bare mass set to zero should be identified
with the $SO(2r)$ Coulomb branch.  This is immediate from the curves
(1.3) and (1.4).
 
\subsec{Sp(2n)}

The corresponding argument for the $Sp(2n)$ curve is less powerful,
since along the flat direction where a single fundamental squark has a
vev, the Higgs mechanism breaking $Sp(2n) \to Sp(2n{-}2)$ only gives
one flavor a mass.  Thus, starting from the scale invariant theory we
flow to a non-asymptotically-free theory at weak coupling on the Higgs
branch.  In order to recover the $Sp(2n{-}2)$ scale invariant theory
we must tune the bare coupling $q{\to}0$ and another bare mass
$m{\to}\infty$ appropriately.  We thus lose any information concerning
the strong--coupling dependence of the original curve. This does not
rule out the possibility that there might be other, more complicated
flat directions that would produce the required $Sp(2n-2)$ curve 
directly.

\bigskip
\centerline{{\bf Acknowledgements}}

It is a pleasure to thank K. Dienes, R. Leigh, R. Plesser, N. Seiberg,
and E. Witten for helpful discussions and comments, and J. Minahan and
D. Nemeschansky for a useful correspondence.
P.C.A. is supported by DOE grant DE-FG05-90ER40559 and
A.D.S. is supported in part by DOE EPSCoR grant DE-FG02-91ER75661 and
by an Alfred P. Sloan Fellowship.

\listrefs
\end